\documentstyle[sprocl]{article}

\input{ epsf}
\def\be{\begin{equation}}
\def\ee{\end{equation}}
\def\bea{\begin{eqnarray}}
\def\eea{\end{eqnarray}}
\begin{document}

\title{COOLING OF NEUTRON STARS REVISITED: APPLICATION OF LOW ENERGY THEOREMS}

\author{A. E. L. DIEPERINK, E. N. E. van DALEN, A. KORCHIN, R. TIMMERMANS}

\address{Kernfysisch Versneller Instituut \\
   Zernikelaan 25, 9747AA Groningen, The Netherlands
    \\ E-mail: dieperink@kvi.nl}

%\author{A. N. OTHER}

%\address{Department of Physics, Theoretical Physics, 1 Keble Road,\\
%Oxford OX1 3NP, England\\E-mail: other@tp.ox.uk}


\maketitle
\abstracts{  Cooling of neutrons stars proceeds mainly via neutrino emission;
  As an example we study the modified neutrino bremsstrahlung process
  $nn \rightarrow nn\nu\bar{\nu}.$
  The radiated energy  is small compared to other scales in the system.
   Hence one can
  use low-energy theorems to compute the neutrino
  emissivity in terms of the non-radiative
  process, ie the on-shell $T$ matrix. We find that the use of the $T-$matrix as
  compared to previous estimates based upon one pion exchange
  leads a substantial reduction of the predicted emissivity. }

\section{Introduction}
 %Motivation\\
  The thermal evolution of neutron stars is dominated by the  weak interaction
and in particular neutrino interactions with the hadronic matter. The hadronic
information is contained in the socalled current-current correlator
$\Pi(q), $ where $q= (\vec{q},\omega)$ is the four-momentum of the neutrinos.
 In general one can distinguish two different regimes:
\\ $\bullet$ neutrino scattering (space-like, $\omega < q )$
 \\ $\bullet$ neutrino-pair emission (time-like, $\omega > q )$.
  \\ In the latter case %(unlike the space-like case)
     the quasi-particle response vanishes, i.e., the one-body process
 $n \rightarrow n+\nu + \bar{\nu} $ is forbidden  by energy-momentum
 conservation, and the socalled URCA process
\\ $\bullet$ $ n \rightarrow p+e^-+ \bar{\nu}  $
\\ \hspace*{1cm} $ p +e^- \rightarrow n+ \nu ,$
\\  is forbidden unless the $n,p$ and $e^-$ fermi momenta satisfy
   the unequality $p_F^p+p_F^e > p_F^n, $
   which  is very unlikely.
\\ Therefore the bremsstrahlung process can only take place in the presence
of spectator   nucleon (referred to as  modified processes), e.g.:
\\ $\bullet$ $n+N\rightarrow n+N +\nu +\bar{\nu} $\ ($N=n,p$)  \\
 $\bullet$ $ n+n  \rightarrow n+p+e^-+ \bar{\nu}  $  (+ inverse)
\\ In the pioneering work of Friman and Maxwell \cite{FM}
these processes were computed in the extreme soft neutrino limit
in Born approximation with the NN interaction
presentated by a Landau type interaction plus a one-pion exchange part.
%in Modelling in past: { In soft-neutrino ($\omega_\nu+\omega_{\bar{\nu}}$
%  one-pion exchange approximation ({ tensor force})
%  to { axial-vector current}.} \\
 \\ With respect to this approach several questions can be asked;
  how accurate is the use of Born approximation, what is the contribution
   of other mesons, and how important are relativistic effects?
 \\ It is the aim of the present study to address these issues by computing
 the emissivity using  a low-energy theorem  on the basis of the observation that
  the neutrino radiation
   is very soft compared to other scales in the system
   ($\omega \approx T= 1$ MeV); this allows one
 to use  an empirical $T-$matrix, fully determined by phase shifts.
 %Other medium effects such as that of quasi-particle width will also de discussed.

%\end{slide*}
%%%%%%%%%%%%%%%%%%%%%%%%%%%%%%%%%%%%%%%%%%%%%%
%\begin{slide*}
%Cooling process proceeds via the { weak interaction}
%\begin{itemize}
%\item{one-body reactions}
%\\ 1)
%\\ 2) $n \rightarrow n+ \nu +\bar{\nu} $   \ { Bremsstrahlung}
% \\ { forbidden by energy-momentum conservation}
% \\ \item{two-body processes}
% \\ 1) $ n+n  \rightarrow n+p+e^-+ \bar{\nu}  $ \ { Modified URCA}
% \\ 2)   $n+n \rightarrow n+n+ \nu +\bar{\nu} $ \ {  Modified BS}
% \\ \item  Radiation is { soft}, i.e. $\omega \sim T \sim 1-10 $MeV
%      { (except for URCA)}
%  \item ``Standard cooling scenarios'' are based upon { one-pion exchange}
%   \end{itemize}
   %\end{slide*}
%%%%%%%%%%%%%%%%%%%%%%%%%%%%%%%%%%%%%%%%%%%%%%%%%%%%%%%

%\begin{slide*}
\section{ Soft electro-weak bremsstrahlung }
First we consider the ``Soft-photon" amplitude in free space \\
The original soft-photon theorem states that the first two terms in an expansion of
the electromagnetic bremsstrahlungs amplitude in photon four-momentum q are fully determined
by the amplitudes of the non-radiative process, \cite{Low}
$$M= A/ \omega+B+ O(\omega) $$
This result can be generalized in several ways, e.g. to virtual bremsstrahlung,
($q^2 >0$) and also to the case of the weak axial vector current.
Here we shall restrict ourselves to the leading order, the $A$ term, in which case
  only radiation off external legs contributes (see Fig. 1), and the amplitude
  can de expressed as
  \begin{figure}
  \epsfxsize=10cm
  \epsfbox{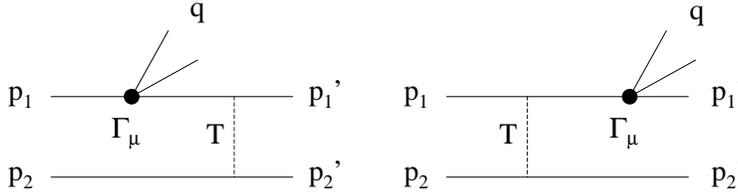}
  \caption{Leading order diagrams for soft bremsstrahlung }
  \end{figure}
$$M_\mu= T { S(p_1-q)}{ \Gamma_\mu} +
  { \Gamma_\mu} { S(p'_1+q)} T +  \{1 \leftrightarrow 2 \} $$
 The weak vector and axial vector vertices $\Gamma$ are given by
 low-energy neutral current Hamiltonian,
 $$H= \frac{G_F}{2\sqrt{2}} B_\mu \bar{\psi} \gamma_\mu (1-\gamma_5) \psi $$
 with the hadron current given by
 $B^\mu = \sum_{i=n,p} \bar{\psi}_i \gamma^\mu (C_{v,i}-\gamma_5 C_{A,i}) \psi_i.$
 Hence in the non-relativistic limit
 $${ \Gamma_\mu(\rm vector)} \approx g_V\gamma_\mu \rightarrow \delta_{\mu,0} $$
  $${ \Gamma^i_\mu(\rm axial)} \approx g^i_A \gamma_5 \gamma_\mu \rightarrow
  g^i_A\vec{\sigma}  \ (i=p,n; \  g_A^p=-g_A^n ) $$
  The two Feynman propagators (corresponding to prior and post emission)
  are given by
  $${ S(p\pm q) }= \frac{1}{\gamma(p \pm q)-m }\approx \frac{\pm m}{p.q}
 = \frac{\pm 1}{\omega- \vec{p}.\vec{q}/m } \approx
 \pm \frac{ 1}{\omega}(1 + O(p.q/m \omega)). $$
  Hence in this order the amplitude
  $M$ can be expressed as a commutator of $T$ with the space component of the
  axial weak current
   operator $$ M_i= \frac{g_A}{\omega} [T, S_i] $$ where
   $S=(\sigma_1+\sigma_2)/2, $
   and in this order
   there is no vector contribution.  \\
  Which terms survive the commutator  $[T,\Gamma^{\rm weak}]$ ?
   Looking at the structure of the NN amplitude given in the appendix
   one sees that there will be non-vanishing contributions from tensor,
   spin-orbit and quadratic spin-orbit terms in $T$. Thus one may conclude
   that only nn spin correlations contribute to the emissivity.
    \begin{figure}
   %\begin{minipage}[t]{6cm}
     \epsfxsize=10cm
  \epsfbox{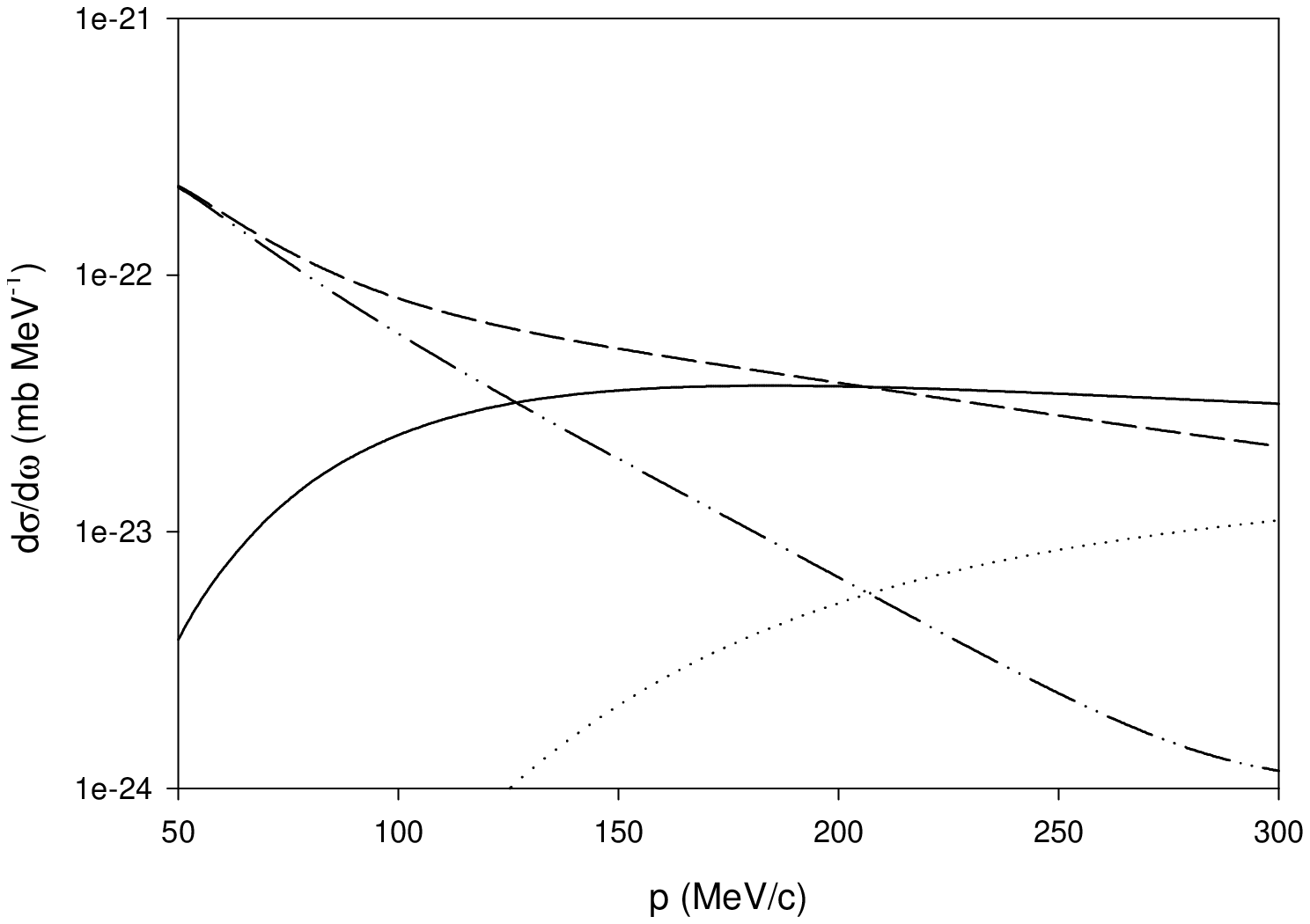}
  %\end{minipage}
  %\begin{minipage}[t]{6cm}
  \epsfxsize=10cm
  \epsfbox{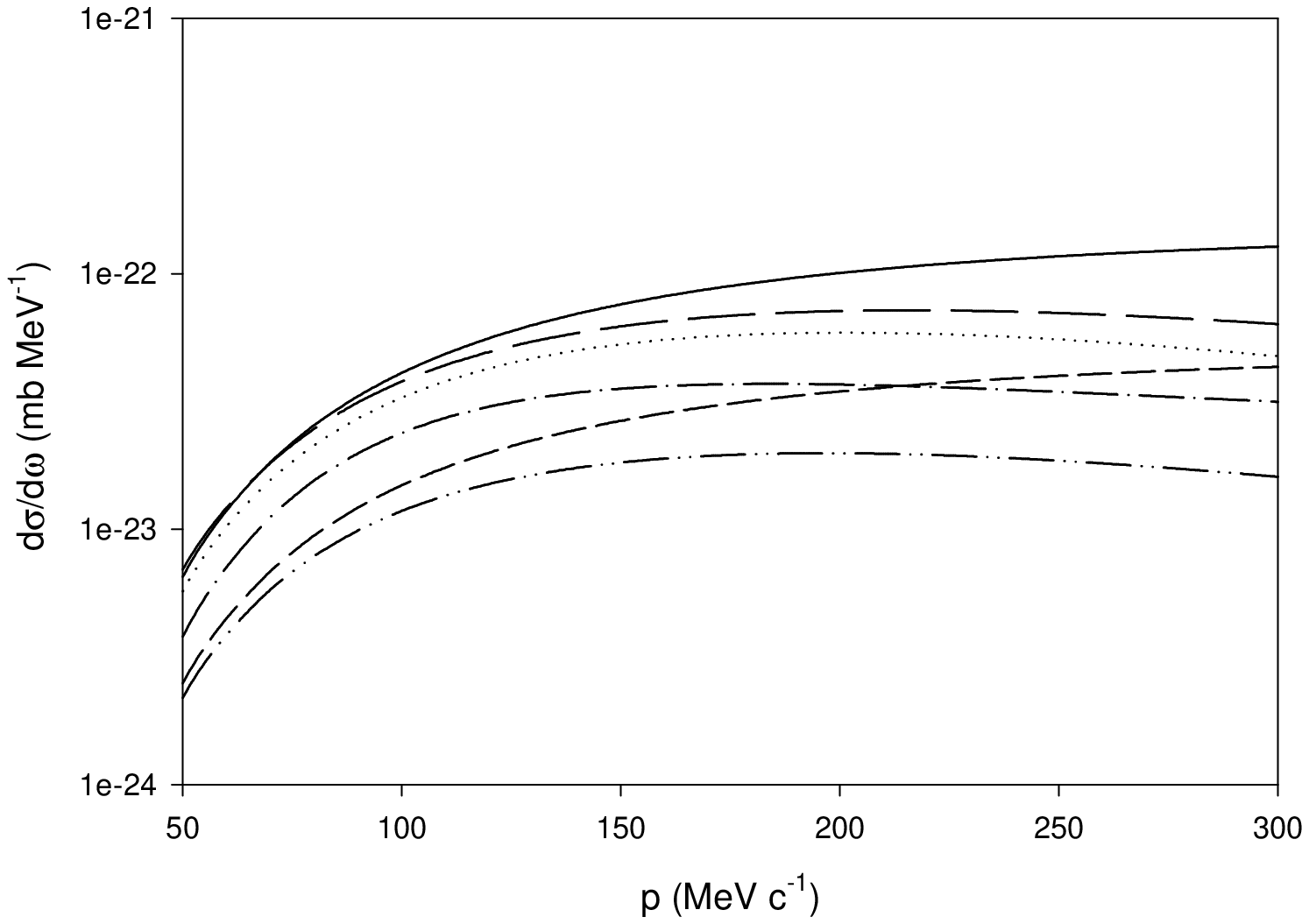}
  %\end{minipage}
  \caption{Cross section for weak nn bremsstrahlung for $\omega$= 1MeV as
   a function of the relative momentum $p=(p_1-p_2)/2; $ top:
    full $T$-matrix (solid line), tensor+spin-spin (dashed), spin-orbit (dotted)
    and quadratic
    spin-orbit (dashed-dotted line); bottom: OPE (solid), OPRE (dotted),
    OPRSE (long dashed), direct OPE (short dashed), direct OPRE
   (dashed-doubly dotted)
    and  full $T$ (dashed-dotted line).  }
       \end{figure}
%\\ Similarly for axial-vector current
%\end{slide*}
%%%%%%%%%%%%%%%%%%%%%%%%%%%%%%%
%nn case \\
%As a result in the extreme soft and non-relativistic limit
% the amplitude can be expressed as
%$$ M_i= \frac{g_A}{\omega} [T, \sigma_i]
  However, we note that in the past
  the effective nn interactions were restricted to purely
local interactions and hence the spin-orbit interactions were not considered.
%{ Spin correlations dominate!}  \\ \\
As an illustation we compute the cross section for NN electro-weak bremsstrahlung
in free space as a function of the relative nn momentum (in the nn cm system).
 In fig. 2a we show the results for full $T-$ matrix and its separate
components. In fig. 2b
the cross section for the full $T-$matrix is compared with the result for
$\pi, \ \pi+\rho $ and also $\sigma$ meson exchange.
\\ We note that the OPE exchange
(which forms the basis of most standard ``cooling scenarios"
overestimates the results obtained with
 the full $T-$matrix
by about a factor 5 for relative momenta in the range appropriate for
neutron matter at normal density ($p_F= 1.3 $fm$^{-1}$)
 (a similar result has been found in \cite{HPR}).
It is also seen that the contribution from the OPE tensor force
 is largely canceled by that from $\rho$
exchange, and that the sum of OPE and ORE is much closer to the T -matrix result
than OPE .
%  \end{slide*}
%%%%%%%%%%%%%%%%%%%%%%%%%%%%%%%%%
\section{The np case} %\begin{slide*} The  np case is slightly more complicated \\
 The np case is slightly different, since the axial vector couplings of neutron and
 proton have opposite values
  $g_A^p= -g_A^n = -1.25. $
  Hence in this case the net result can be expressed as the sum of a commutator
  and an anticommutator, where the latter corresponds to the exchange diagrams
  $$M_i = \frac{g_A}{\omega} ([T^{dir}, D_i] + \{T^{exch}, D_i\}) $$
  where $D= (\sigma_1- \sigma_2)/2, $ and $T^{dir,exch}$ are defined in the appendix.
\\ %\end{slide*}
In this case also the socalled Landau terms in the $T$-matrix, of the form
$ g\sigma_1.\sigma_2 $and
$ g'\sigma_1.\sigma_2 \tau_1.\tau_2 $ contribute  to the anti-commutator.
In practice the proton fraction in a neutron star is quite small and
therefore we find that the relative contribution from np is quite unimportant.
\section{Emissivity in medium}
%%%%%%%%%%%%%%%%%%%%%%%%%%%%%%%%%%
%{\bf Emissivity in medium} \\
The emissivity in the medium with given density and temperature $T$
can be computed from the amplitude
in two ways: (i) via the (imaginary part) of the current-current correlator,
 or (ii) using the Fermi
golden rule. The former approach is more general, but the latter is simpler to use
and is sufficient in the present case.
In lowest order in the virial expansion
it amounts to convoluting  the free space matrix
elements $|M^2|$ with the hadronic
Fermi-Dirac functions
$$ \epsilon =  \Pi_i \int d^3p_i \frac{d^4q_1}{\omega_1} \frac{d^4q_2}{\omega_2}
(\omega_1+\omega_2) \delta^4(p_1+p_2-p_3-p_4-q_1-q_2) F |<M_{j}>|^2 ,  $$
where $F= f_1f_2 (1-f_3)(1-f_4).$
In practice the integrals are simplified by taking
the absolute values of the nucleon momenta equal to the fermi momenta.
If we consider the ratio of the emissivity calculated with the full T matrix
and the one from the OPE \cite{FM} we obtain about a factor 4-5 reduction.
This is inline with the conclusion in \cite{HPR} for pure neutron matter.
\\ Other medium that need to be considered in
further work are 1) replacement of the free $T$- by a $G$-matrix,
which takes into account the Pauli blocking and other medium effects,
2) medium renormalization of the axial coupling vertex,
3) inclusion of higher order medium effects such as dressing Green functions.
In the latter case one has to take care to conserve the symmetries (CVC, PCAC)
of the problem (which are conserved in the present case).

\section*{Acknowledgments}
We thank A. Sedrakian for stimulating discussions. This work has been supported
by the Stichting voor Fundamenteel Onderzoek der Materie with financial
support from the Nederlandse Organisatie voor Wetenschappelijk Onderzoek.
%This is where one places acknowledgments for funding
%bodies etc.  Note that there are no section numbers for
%the Acknowledgments, Appendix or References.

\section*{Appendix: Structure of the on-shell NN amplitude}
The on-shell T-matrix is fully determined by the nn phaseshifts.
First we consider nn; in a covariant approach one has
$$T^{on}= T_\alpha(s,t,u)\bar{u}(p_2)\Omega_\alpha u(p_2) \times
\bar{u}(p_1) \Omega_\alpha u(p_1) $$
with
$$ \Omega_\alpha =
\{ 1,\sigma_{\mu,\nu}, \gamma_\mu, \gamma_5, \gamma_\mu \gamma_5 \}$$
Non-relativistically
$$T=T_c + T_s\sigma_1.\sigma_2 +T_T S_{12} +T_{SO} L.S +T_Q Q_{12} $$
 In case of np there are twice as many components which can be distinguished
 by isospin $I=0,1$ , i.e. $ T^{dir}= ( T^0+T^1 )/2 $ and
  $ T^{exch}= (T^0- T^1)/2. $
  % \\ \\ \\ The Spin response function for $q \rightarrow 0$
%$$S_{sp} (\omega) \equiv \Im \Pi(\omega)
%  = \Im \int dt e^{i\omega t} < \sigma(q,t) \sigma(q,0)> $$
% $$ = \int \Pi_i d^3p_i \delta(p_1+p_2-p_3-p_4) F |<M_{i}>|^2  $$
% $$ \epsilon=  \int d \omega \omega^6 S_{sp}(\omega) $$
%axion emission  $\Gamma(axion)= g_{ann}\gamma_5 $ \\
%photon emission  $\Gamma(em) = e\gamma_\mu $
%%%%%%%%%%%%%%%%%%%%%%%%%%%%%%%%%%%
%\begin{slide*}
%Emissivity
%\end{slide*}

\end{document}